# Universal Machine Learning Kohn-Sham Hamiltonian for Materials

Yang Zhong[1,2], Hongyu Yu[1,2], Jihui Yang[1,2], Xingyu Guo[1,2], Hongjun Xiang[1,2,*], and Xingao Gong[1,2]

[1]*Key Laboratory of Computational Physical Sciences (Ministry of Education), Institute of Computational Physical Sciences, State Key Laboratory of Surface Physics, and Department of Physics, Fudan University, Shanghai, 200433, China*
[2]*Shanghai Qi Zhi Institute, Shanghai, 200030, China*

Email: hxiang@fudan.edu,cn

## Abstract

While density functional theory (DFT) serves as a prevalent computational approach in electronic structure calculations, its computational demands and scalability limitations persist. Recently, leveraging neural networks to parameterize the Kohn-Sham DFT Hamiltonian has emerged as a promising avenue for accelerating electronic structure computations. Despite advancements, challenges such as the necessity for computing extensive DFT training data to explore each new system and the complexity of establishing accurate ML models for multi-elemental materials still exist. Addressing these hurdles, this study introduces a universal electronic Hamiltonian model trained on Hamiltonian matrices obtained from first-principles DFT calculations of nearly all crystal structures on the Materials Project. We demonstrate its generality in predicting electronic structures across the whole periodic table, including complex multi-elemental systems, solid-state electrolytes, Moiré twisted bilayer heterostructure, and metal-organic frameworks (MOFs). Moreover, we utilize the universal model to conduct high-throughput calculations of electronic structures for crystals in GeNOME datasets, identifying 3,940 crystals with direct band gaps and 5,109 crystals with flat bands. By offering a reliable efficient framework for computing electronic properties, this universal Hamiltonian model lays the groundwork for advancements in diverse fields, such as easily providing a huge data set of electronic structures and also making the materials design across the whole periodic table possible.

## Introduction

The electronic structure[1-5] of materials is crucial in understanding and predicting a wide range of physical properties, including electrical conductivity, optical behavior, mechanical strength, chemical reactivity, and magnetic characteristics. Electronic structure calculations provide insights into the electronic band structures, bonding, and reactivity, enabling the design of new materials and the study of chemical reactions.

Among diverse quantum mechanics approaches, density functional theory (DFT)[6-9] has become a widely used computational method in electronic structure calculations since DFT has drastically reduced the computational cost by employing the electron density instead of the many-body wave function as the fundamental variable. The price to pay for using the electron density as the fundamental variable is that the Kohn-Sham DFT Hamiltonian is no longer a simple explicit function of the atomic structure, but should be obtained by solving a self-consistent Kohn-Sham equation. However, the computational cost of DFT self-consistent field (SCF) cycles remains expensive for large systems.

In recent years, the use of neural networks to parameterize the DFT Hamiltonian has emerged as an important and effective method to accelerate electronic structure calculations[10-22]. The ML Hamiltonian models offer a significant advantage by providing a direct mapping from the structure to the self-consistent Hamiltonian matrix, eliminating the need for time-consuming self-consistent iterations typically required in Kohn-Sham DFT. Initially, Hegde and Bowen proposed a kernel ridge regression (KRR) model and successfully applied it to fit the empirical Hamiltonian matrix of the cubic Cu crystal[10]. Subsequently, Unke[12] proposed the PhiSNet model, and Schütt[14] introduced the SchNOrb model, both of which demonstrated remarkable performance in accurately fitting the Hamiltonian matrices of various small organic molecules such as water, ethanol, and uracil. In addition, some other researchers have also made important contributions. For instance, Nigam[18] used the Gaussian Process Regression (GPR) model to fit the Hamiltonian matrices of water and benzene molecules, while Zhang[20] successfully fitted the Hamiltonian matrix of aluminum using the atomic cluster expansion (ACE) model. Xu and coworkers proposed the graph neural network (GNN)-based DeepH[17, 21] model, which was used to predict the Hamiltonian matrices of crystals like graphene and $MoS_2$.

However, despite these advancements, there still exist several challenges in utilizing machine learning for electronic Hamiltonian prediction. Firstly, exploring new systems necessitates retraining a completely new model specifically tailored for that system, which currently lacks automation and can prove time-consuming and computationally expensive, thereby constraining the practical applicability of this approach. Secondly, many practical materials, such as high-entropy alloys[23-25] and ceramics[26, 27], are composed of a large number of elements, posing challenges in establishing accurate machine-learning models for their electronic Hamiltonians due to the requirement of a substantial amount of DFT training data. Current research endeavors have predominantly concentrated on systems comprising no more than three elements, and surmounting this obstacle to encompass more intricate systems remains a significant challenge in the field. Overcoming these challenges is crucial to enable the broader application of machine learning in predicting electronic Hamiltonians for diverse and multi-elemental materials. Recently, several groups reported developments of universal machine learning interatomic potentials (MLIPs)[28-33], which can handle almost all elements across the whole periodic table. Naturally, one might wonder whether it is possible to develop a universal machine-learning model for electronic Hamiltonians. Such a model would enable efficient and accurate calculations of electronic structures for a wide range of materials and large-scale systems. Recently, we have proposed a transferable Hamiltonian graph neural network (HamGNN)[22] that enables the prediction of Hamiltonian matrices for various structures within a chemical space comprising several specific elements, such as $SiO_2$ isomers and $Bi_xSe_y$ compounds with varying stoichiometric ratios. Compared to the above mentioned models, the excellent transferability of the equivariant GNN-based HamGNN framework makes it the most feasible candidate for constructing a universal Hamiltonian model across the entire periodic table. Unlike the construction of universal MLIPs, achieving a universal Hamiltonian model is not as straightforward and cannot simply rely on training with large-scale datasets due to the high dimensionality and inherent complexity of the Hamiltonian matrix. The training process also plays a crucial role in achieving the universality of the Hamiltonian model.

In this work, we propose a universal Kohn-Sham Hamiltonian model in the sense that this single model is applicable to all elements of the entire periodic table and all structures of a given chemical composition. By employing the HamGNN model, we have developed a universal Hamiltonian model via a 'two-step training procedure' utilizing the Hamiltonian matrices of 5,5000 structures obtained from the Materials Project[34, 35] as our training dataset. The universality of our Hamiltonian model is demonstrated by its successful prediction of the electronic structures for various bulk or low-dimensional materials with different combinations of chemical elements in the periodic table. The universal model successfully captures not only common systems but also those containing uncommon or rare transition metal elements. Furthermore, it proves high accuracy even in complex multi-element systems comprising more than five elements. By providing a reliable framework for understanding electronic properties across the periodic table, this research paves the way for advancements in material design, catalysis, electronics, and other fields that heavily rely on efficient predictions of electronic structures.

## Results

### *The framework of the universal electronic Hamiltonian model*

Constructing an ML model for the electronic Hamiltonian is more complicated than constructing machine learning interatomic potentials (MLIPs)[36-41]. MLIPs provide a mapping from the two degrees of freedom, atomic types $\{Z_i\}$ and atomic positions $\{\tau_i\}$, to the scalar potential energy $E$. In addition to the two degrees of freedom, the mapping from a crystal structure to the electronic Hamiltonian matrix also necessitates handling the supplementary degree of freedom arising from distinct atomic orbital bases $\{\phi_{i\alpha}\}$ associated with each atom. As the number of elements in the system increases, the relevant degrees of freedom and interactions in the electronic Hamiltonian matrix also increase dramatically, leading to a significant increase in complexity when fitting the electronic Hamiltonian matrix. Therefore, a universal electronic Hamiltonian model across the periodic table requires much more network capacity than the MLIPs to capture all degrees of freedom accurately. For large molecules or crystals, especially in cases where the basis set is large and the system is complex, the dimension of the electronic Hamiltonian matrix can be extremely large. This increases the difficulty of model training, as it requires handling large-scale matrix and storing a significant amount of parameters. In addition, different sub-blocks of a Hamiltonian matrix are subject to different equivariant constraints under a rotation operation, so the Hamiltonian matrix predicted by the model must also adhere to these constraints.

We have trained a universal Hamiltonian model following the process shown in Figure 1. To develop a universal Hamiltonian model for the whole periodic table, we utilized one of the world's largest open databases for DFT-relaxed crystal structures, namely the Materials Project[34, 35]. We used OpenMX[42, 43], a DFT software package based on norm-conserving pseudopotentials and pseudo-atomic localized basis functions, to calculate the Hamiltonian matrices of approximately 55,000 structures on the Materials Project. Among them, approximately 44,000 structures' Hamiltonian matrices were used as a training set, while validation and test sets were created using the Hamiltonian matrices of around 5,500 structures each. Then, we use these datasets to train a universal HamGNN[22] model. HamGNN is a deep learning model based on equivariant graph neural networks, which can automatically learn the features of each element on the entire periodic table without any prior physical or chemical properties of elements. The architecture of the HamGNN model and the training process are shown in Figure 1(b). We will briefly introduce the principle of HamGNN for predicting the Hamiltonian matrix in the following paragraph, and for more network details see Ref. 22.

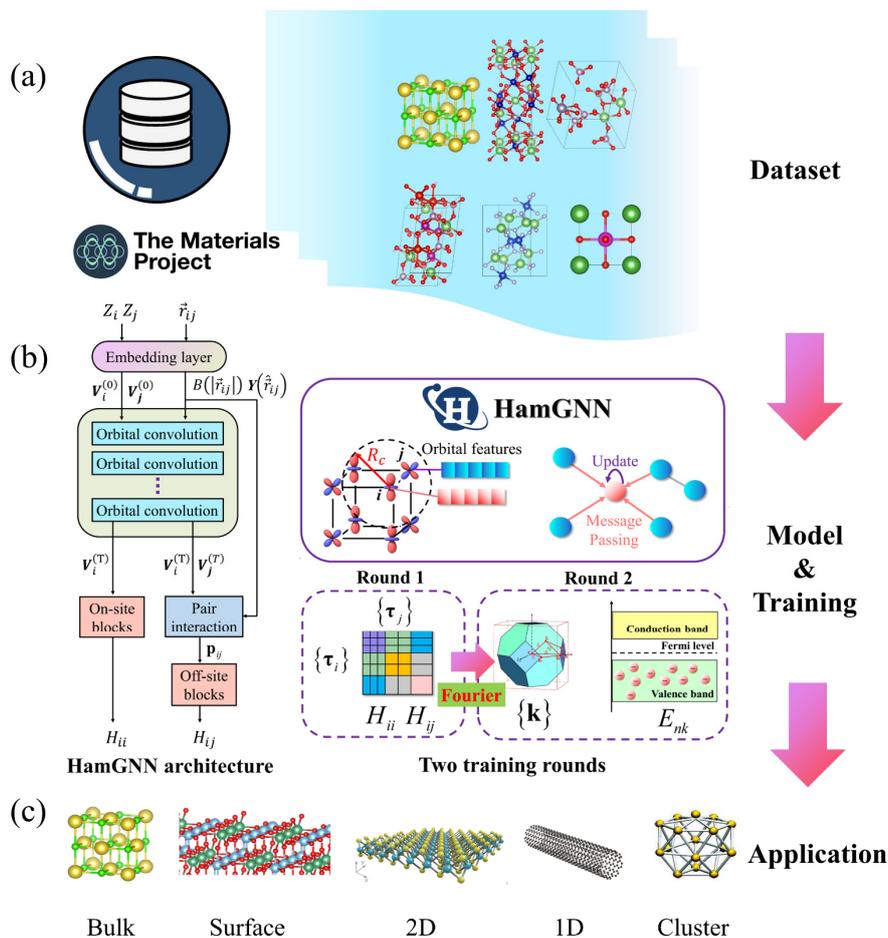

**Figure 1**. The framework of a universal Hamiltonian model based on HamGNN. (a) Training dataset preparation. The training dataset is generated by calculating the real-space Hamiltonian matrices of crystal structures available on the Materials Project using an ab initio tight-binding software based on numerical atomic orbitals. (b) Model architecture and the training procedure. This dataset is utilized for training the HamGNN model, a deep learning approach that employs equivariant graph neural networks to predict the Hamiltonian matrix. The model can automatically learn the intrinsic features of each element on the periodic table solely based on their atomic numbers, without relying on any prior physical or chemical properties. In HamGNN, the orbital features of the central atom are updated by considering interactions between the neighbor atoms within a cutoff radius of $R_c$. For atomic pairs beyond this cutoff radius, multi-layer message passing is employed to exchange orbital features. To achieve universality, HamGNN requires two rounds of training. In the first round, the loss function for network training solely considers the error in the real-space Hamiltonian matrix. After this initial round of training, this model can accurately predict the real-space Hamiltonian matrices with high precision. In the second round of training, the real-space Hamiltonian matrices are transformed into the reciprocal Hamiltonian matrices at randomly selected **k** points in the Brillouin zone, and the errors of the orbital energies near the Fermi level obtained by diagonalizing the reciprocal Hamiltonian matrices are incorporated into the total loss function. (c) Applications of the universal HamGNN model. After two rounds of training, the universality of the HamGNN model has significantly improved, enabling accurate prediction of electronic structures in crystals with arbitrary periodic boundary conditions and any components.

Since the irreducible representations with rotation orders $l$ = 0, 1, 2... of the O(3) group possess the same rotational equivariance and parity symmetry as the atomic orbitals $s$, $p$, $d$..., HamGNN uses a direct sum of equivariant atomic features $\boldsymbol{V}_l$ with different rotation orders $l$ to characterize each atom: $\boldsymbol{V} = \boldsymbol{V}_0 \oplus \boldsymbol{V}_1 \oplus \cdots \oplus \boldsymbol{V}_{l_{max}}$. This feature tensor satisfies the rotational equivariance under the rotation operation $\hat{R}$: $\boldsymbol{V}_l\left(\hat{R} \cdot (\vec{r}_1, \cdots, \vec{r}_N)\right) = \boldsymbol{D}_l(\hat{R})\boldsymbol{V}_l(\vec{r}_1, \cdots, \vec{r}_N)$, where $\boldsymbol{D}_l$ ($l < l_{max}$) is a Wigner $D$ matrix[44, 45] of order $l$. HamGNN updates the equivariant atomic features through an equivariant message-passing function in the orbital convolution layer. After $T$ orbital convolution layers, the atomic features are transformed into on-site Hamiltonian matrices by an "on-site layer". HamGNN merges the features of atom pairs $ij$ into the edge features $\boldsymbol{P}_{ij}$ in the "Pair interaction layer". The edge features $\boldsymbol{P}_{ij}$ is later transformed into an off-site Hamiltonian matrix through an "off-site layer". Each subblock of the Hamiltonian matrix can be decomposed into a set of O(3) equivariant irreducible spherical tensors (ISTs) according to the following equation[44-47]:

$$l_i \otimes l_j = |l_i - l_j| \oplus |l_i - l_j| + 1 \oplus \cdots \oplus l_i + l_j \tag{1}$$

The on-site and off-site layers output each sub-block of the Hamiltonian matrix by the following equation

$$H_{l_i m_i, l_j m_j} = \sum_{l=|l_i-l_j|}^{l_i+l_j} \sum_{m=-l}^{l} C_{m_i,m_j,m}^{l_i,l_j,l} T_m^l \tag{2}$$

where $T_m^l$ is an equivariant IST with rotation order $l$ in $\boldsymbol{V}_i^{(T)}$ or $\boldsymbol{P}_{ij}$, $C_{m_i,m_j,m}^{l_i,l_j,l}$ is the Clebsch-Gordan coefficient.

To achieve a universal model, HamGNN needs to undergo two rounds of training. During the first round of training, only the error of the real-space Hamiltonian matrix is considered as the loss function for the network's training. After the first round of training, the model is capable of reasonably predicting the real-space Hamiltonian matrix, which lays a good foundation for obtaining accurate band structures in subsequent tasks. To improve the accuracy of the model in predicting the eigenvalues of Bloch states, which constitutes the primary objective of our Hamiltonian model, we further applied fine-tuning techniques for a second round of training. Fine-tuning in neural networks has shown a crucial role in training recently developed large language models to optimize their adaptation to specific tasks or domains[48, 49]. The results of our tests indicate that the two training rounds are necessary to obtain a truly universal Hamiltonian model. During each training step in the second round, Fourier transformations are performed on the predicted and target real-space Hamiltonian matrices at randomly selected **k** points in the Brillouin zone. The orbital energies $\varepsilon_{nk}$ are obtained by diagonalizing the reciprocal Hamiltonian matrices, and the error of the orbital energies near the Fermi level is incorporated into the loss function as follows:

$$L = \|\tilde{H} - H\| + \frac{\lambda}{N_{orb} \times N_k} \sum_{k=1}^{N_k} \sum_{n=1}^{N_{orb}} \|\tilde{\varepsilon}_{nk} - \varepsilon_{nk}\| \tag{3}$$

where the variables marked with a tilde refer to the corresponding predictions and $\lambda$ denotes the loss weight of the orbital energy error. $N_{orb}$ is the number of orbits selected near the Fermi level, $N_k$ is the number of the random **k** points generated in each training step.

Figure 2(a) displays the count of each chemical element in the training set for the Hamiltonian. The metallic elements of Group IA and IIA, as well as the non-metallic elements of Group IVA, VA, VIA, and VIIA, have the highest proportion in the training set. Except for some less common transition metal elements that are not included in the training set, this dataset includes all elements supported by OpenMX's PBE pseudopotential library. After the first round of training, the HamGNN model achieved a mean absolute error (MAE) of only 5.4 meV for the real-space Hamiltonian matrix on the test set. The accuracy of the model for the real-space Hamiltonian matrix is shown in Figure 2(b). In the second round of training, we incorporate the error of orbital energies at five randomly selected **k**-points in the Brillouin zone into the loss function to restart the training, with a $\lambda$ value set at 0.01. After the second round of model fine-tuning, the generalization ability of the HamGNN model has significantly improved. The model's high accuracy on the test set is evident from its predictions of energy bands and Fermi surfaces for several crystal structures in the test set (see Supplementary Discussion 1). By addressing potential overfitting issues, this model demonstrates better adaptability to different datasets and real-world scenarios, showcasing enhanced universality and stability. In the subsequent discussions, we will evaluate HamGNN's prediction accuracy for more complex crystals across the entire periodic table.

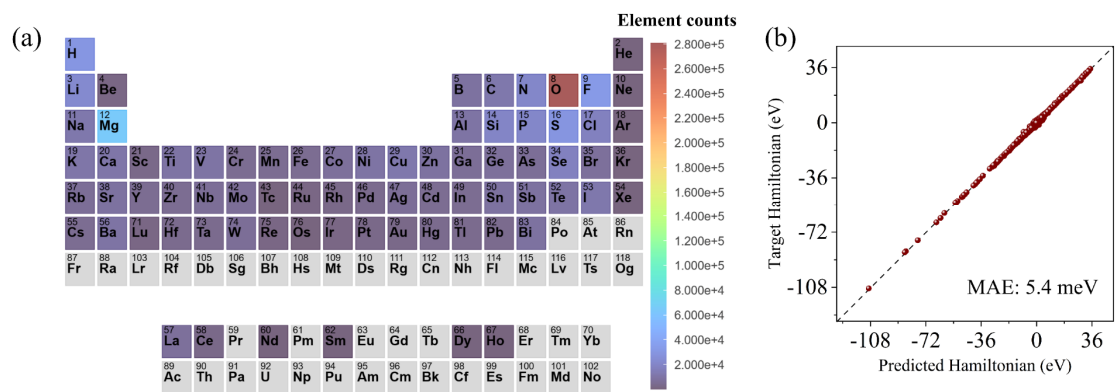

**Figure 2**. (a) The element distribution of the training dataset. (b) The comparison of the Hamiltonian calculated by HamGNN and OpenMX for the test dataset.

### *Tests on multi-element materials*

We are now applying our universal Hamiltonian model to systems that are not included in the Materials Project to verify the generality and accuracy of our model. Previous ML Hamiltonian models[10, 15, 20] typically handle crystal structures composed of only 1 to 3 elements, and training an accurate model that can deal with complex crystal structures containing more elements is challenging. The training datasets commonly used by these models are built from structures perturbed using molecular dynamics. However, as the number of atomic species in the crystal increases, the degrees of freedom of the Hamiltonian matrix increase sharply. Consequently, the training samples generated by perturbing a seed structure cannot fully cover the entire configuration space of the crystal. In these cases, the perturbed structures often contain many similar and repetitive patterns, which can cause the Hamiltonian model to be trapped in local minima and fail to accurately fit the Hamiltonian of crystals with multiple elements and complex configurations. However, the universal Hamiltonian model can effectively address such concerns. Through extensive training on a comprehensive and diverse dataset, the universal Hamiltonian model develops a profound understanding of the intricate interactions among atoms in various configurations.

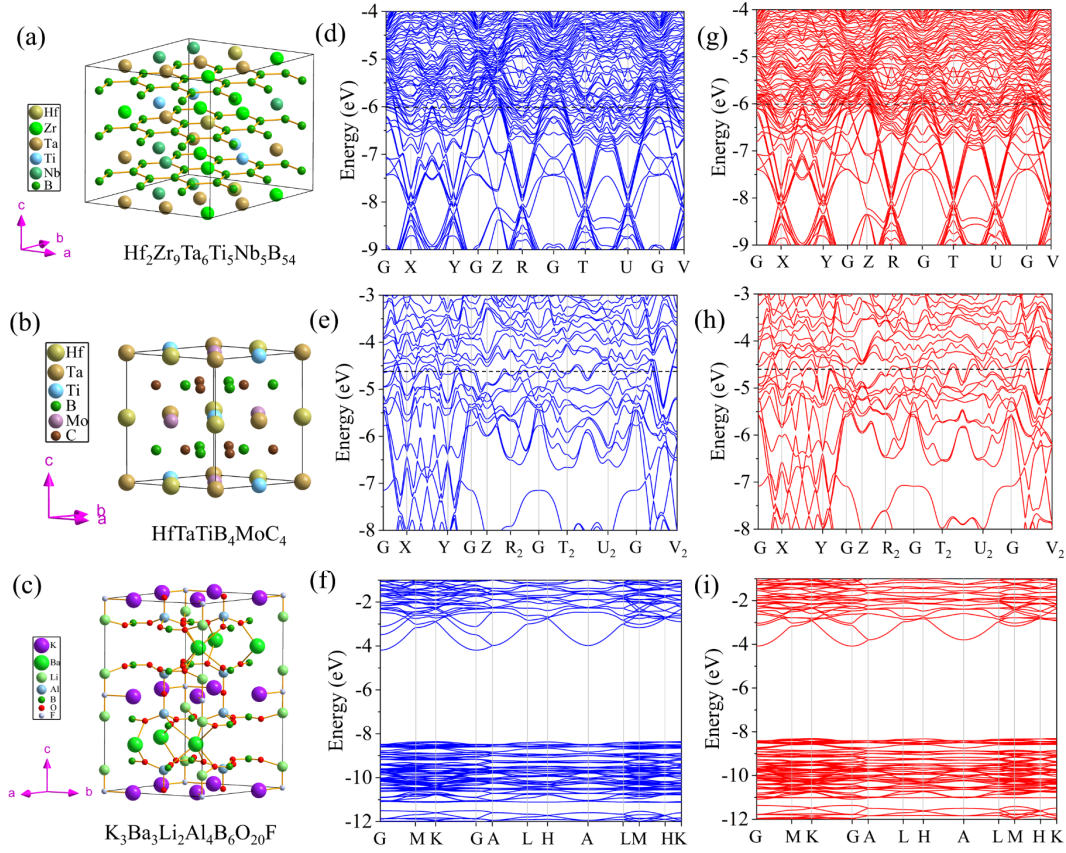

**Figure 3**. The crystal structures of (a) $Hf_2Zr_9Ta_6Ti_5Nb_5B_{54}$, (b) $HfTaTiB_4MoC_4$, and (c) $K_3Ba_3Li_2Al_4B_6O_{20}F$. The predicted energy bands of (d) $Hf_2Zr_9Ta_6Ti_5Nb_5B_{54}$, (e) $HfTaTiB_4MoC_4$, and (f) $K_3Ba_3Li_2Al_4B_6O_{20}F$. The DFT calculated energy bands of (g) $Hf_2Zr_9Ta_6Ti_5Nb_5B_{54}$, (h) $HfTaTiB_4MoC_4$, and (i) $K_3Ba_3Li_2Al_4B_6O_{20}F$.

The universal HamGNN's generalization performance and accuracy were evaluated by conducting tests on three different crystal structures: $Hf_2Zr_9Ta_6Ti_5Nb_5B_{54}$, $HfTaTiB_4MoC_4$, and $K_3Ba_3Li_2Al_4B_6O_{20}F$. These crystals were specifically chosen to represent a diverse range of compositions and structural complexities[50-52]. The compound $Hf_2Zr_9Ta_6Ti_5Nb_5B_{54}$ exhibits a hexagonal $\omega$-phase derived structure and belongs to the *P*1 space group of the triclinic crystal system. It consists of five distinct metal ions, which are inserted randomly into the gaps between the hexagonal monolayers of boron. $HfTaTiB_4MoC_4$ crystallizes in the triclinic *P*1 space group, with the metal elements inserted into the gaps between the hexagonal monolayers composed of boron and carbon. $K_3Ba_3Li_2Al_4B_6O_{20}F$ is a deep ultraviolet transparent nonlinear optical crystal, composed of seven elements and possessing a large bandgap[53].

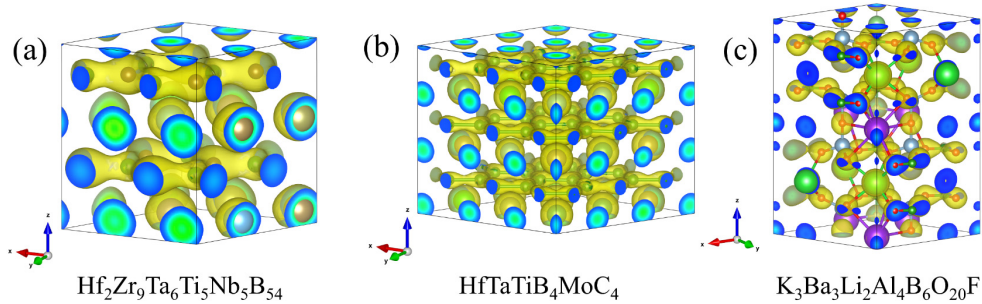

**Figure 4**. The predicted charge density of (a) $Hf_2Zr_9Ta_6Ti_5Nb_5B_{54}$, (b) $HfTaTiB_4MoC_4$, and (c) $K_3Ba_3Li_2Al_4B_6O_{20}F$.

To assess the generalization performance of HamGNN, a comparison was made between the predicted energy bands generated by the model and the calculated energy bands by OpenMX for each of these crystals, as shown in Figure 3. By examining Figure 3, it becomes evident that HamGNN demonstrates remarkable generalizability in predicting the energy bands for the three crystal structures. The charge densities of these three crystals were also obtained using the predicted Hamiltonian matrix, as shown in Figure 4. The predicted charge densities exhibit excellent agreement with the corresponding DFT-calculated charge densities (shown in Supplementary Figure. S6). This further confirms that HamGNN is capable of accurately capturing not only energy bands but also charge distribution within these materials. Furthermore, to test its versatility and applicability on practical materials, we extended our analysis to a sulfide solid electrolyte with a composition of $Li_{10}Si_{1.5}P_{1.5}S_{11.5}Cl_{0.5}$, a highly disordered system consisting of 7200 atoms. Our universal model shows that this disordered system is insulating with a band gap of 1.4 eV, suggesting that it indeed has excellent electric insulating properties and may serve as a promising solid electrolyte (see Supplementary Discussion 2 for details). These successful evaluations highlight the effectiveness and reliability of HamGNN as a universal model for predicting electronic structures across various complex crystal systems. Its ability to generalize well across different compositions and structural complexities makes it an invaluable tool in high-throughput electronic structure calculations for the periodic table.

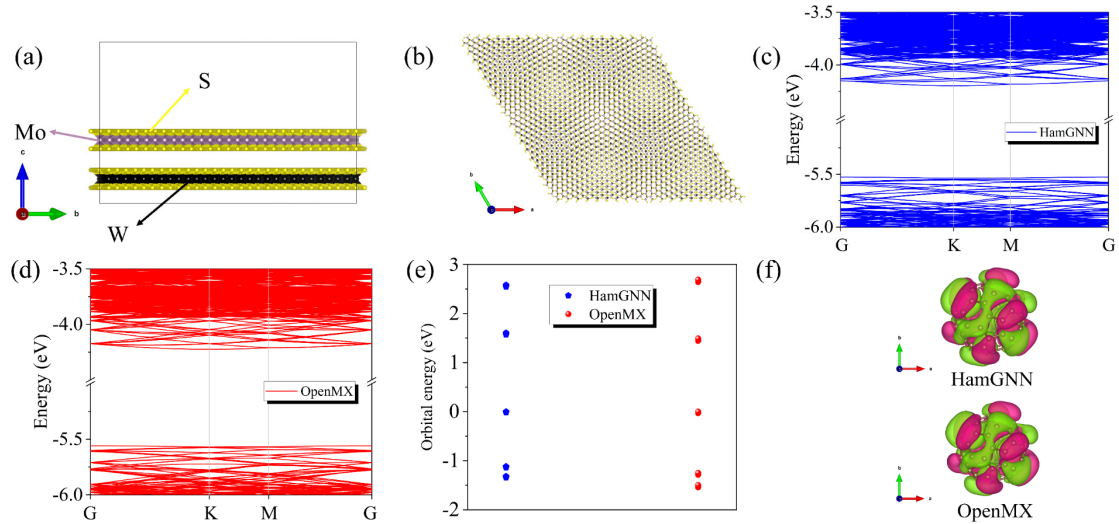

**Figure 5**. Prediction results of the universal HamGNN model for the bilayer $MoS_2/WS_2$ heterostructure with a twist angle of 3.5° and the C60 cluster. (a) Side view of the bilayer $MoS_2/WS_2$ heterostructure. (b) Top view of the bilayer $MoS_2/WS_2$ heterostructure. (c) Predicted band structure of the bilayer $MoS_2/WS_2$ heterostructure. (d) Band structure of the bilayer $MoS_2/WS_2$ heterostructure calculated by OpenMX. (e) The HamGNN-predicted and OpenMX-calculated orbital energies for the C60 cluster. (f) The HamGNN-predicted and OpenMX-calculated wavefunction for the highest occupied molecular orbital (HOMO) of the C60 cluster.

*Tests on low-dimensional materials*

The above discussions have demonstrated the accuracy of the universal HamGNN model on bulk materials. Now, we will further explore its generalizability and prediction accuracy in the field of low-dimensional materials. We constructed a two-dimensional heterostructure consisting of $MoS_2/WS_2$ with a twist angle of 3.5°, as shown in Figure 5(a) and Figure 5(b). This structure comprises 1625 atoms and exhibits a higher level of complexity compared to the bulk structures typically included in our training set. The universal HamGNN model effectively captures the interatomic interactions within the bilayer $MoS_2/WS_2$ heterostructure and demonstrates excellent agreement between the energy bands predicted by the universal HamGNN model (Figure 5(c)) and those calculated by OpenMX (Figure 5(d)). The universal HamGNN model was further tested on the C60 cluster. Figure 5(e) demonstrates a good alignment between the predicted energy level of the C60 cluster near the band gap and the results obtained through DFT calculations, while Figure 5(f) illustrates a close match between the predicted wave function and the wave function calculated by DFT. These tests demonstrate the powerful and wide applicability of the universal HamGNN model in various material systems, ranging from bulk crystals to low-dimensional materials. By utilizing this universal model, researchers can explore a vast array of low-dimensional materials with tailored functionalities and properties.

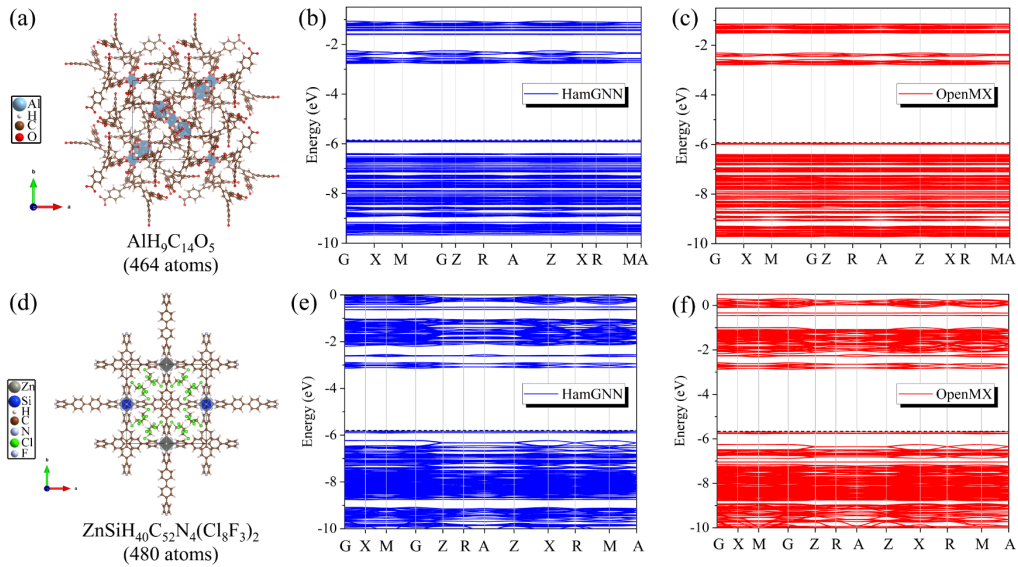

**Figure 6**. (a) The crystal structure, (b) predicted energy bands, and (c) calculated energy bands of AlH$_9$C$_{14}$O$_5$. (d) The crystal structure, (e) predicted energy bands, and (f) calculated energy bands of ZnSiH$_{40}$C$_{52}$N$_4$(Cl$_8$F$_3$)$_2$.

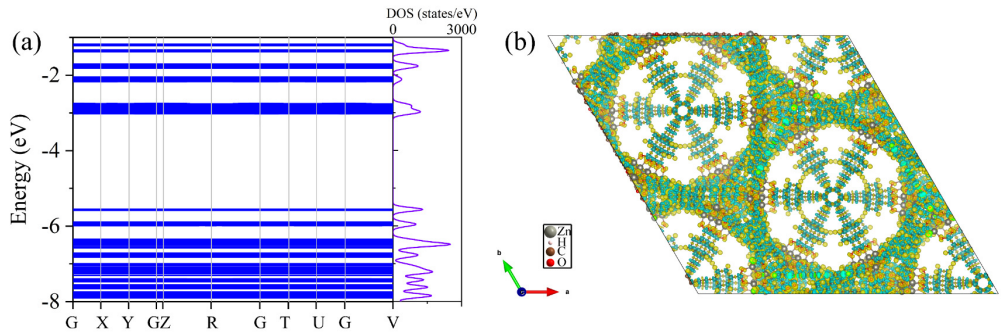

**Figure 7.** (a) The predicted energy bands and (b) the difference charge density of Zn$_4$H$_{28}$C$_{58}$O$_{13}$. The yellow and blue colors represent charge accumulation and depletion, respectively.

## *Application of the universal Hamiltonian model to large-scale hybrid inorganic-organic crystals*

In this section, we will utilize the universal Hamiltonian model to predict the electronic structures of metal-organic frameworks (MOFs)[54-56] and demonstrate its efficiency and broad applicability in studying hybrid inorganic-organic crystals. MOFs[54-56] are a type of porous material composed of metal ions and organic ligands, forming intricate three-dimensional network structures. The electronic structures[57, 58] of MOF materials, such as band structure, electron density of states, and electron orbital distribution, directly influence the conductivity, optical properties, and potential applications in fields like photocatalysis and photovoltaics. However, due to the complex structure and large size of MOFs, accurately predicting their electronic structures using DFT can be challenging and computationally expensive.

By training on inorganic crystal structures available on the Materials Project, we have successfully developed a universal Hamiltonian neural network model that demonstrates high accuracy across various types of inorganic materials. However, this model lacks training on organic crystal structures which encompass a significant number of covalent bonds. Consequently, when predicting some complex organic or hybrid inorganic-organic crystal structures, the model may encounter certain challenges. In order to further improve the accuracy of the model in predicting the Hamiltonian matrix of organic crystal structures, we employed incremental training to further train and fine-tune the model. We selected approximately only 1800 small MOF crystal structures from the QMOF[59] Database, with a maximum number of atoms per unit cell not exceeding 50, as the training set. Subsequently, we further restart the training of the previous universal HamGNN model using this dataset. The mean absolute error of the Hamiltonian matrix predicted by the further trained model on this MOF dataset is about 3.95 meV. The comparison between the Hamiltonian matrix elements predicted by HamGNN and those calculated by OpenMX on the small MOF dataset is shown in Supplementary Figure S8. After undergoing fine-tuning, the Hamiltonian model demonstrates enhanced precision in inferring the interactions among diverse covalent bonds within organic crystalline materials.

As a test, we used the model to predict the electronic structures of two MOF materials, $AlH_9C_{14}O_5$ and $ZnSiH_{40}C_{52}N_4(Cl_8F_3)_2$, and compared them with the results obtained from DFT calculations. Both crystal structures have complex topological configurations, and the latter even contains up to seven elements, posing great challenges to the model. Despite these complexities, the energy bands predicted by the model for both structures are in good agreement with those obtained from DFT calculations, as shown in Figure 6. This indicates that our developed model has high accuracy and reliability when dealing with complex organic crystal structures. Furthermore, the difference charge densities predicted for these two materials achieved excellent consistency with those obtained from DFT calculations, as shown in Supplementary Figure S9. Moreover, the speed of DFT calculation can be improved by several orders of magnitude by using machine learning methods. Taking $ZnSiH_{40}C_{52}N_4(Cl_8F_3)_2$ as an example, the time cost of using DFT methods is as high as 181 core·hours. However, the HamGNN model only takes 0.33 core·hours to complete the computational task.

With further research, some more complex and larger MOF materials show wide application prospects in catalyst design, photovoltaics, solar cells, light-emitting diodes (LEDs), etc[60]. In this case, the advantage of using machine learning Hamiltonian models for computing large MOF materials becomes apparent. To demonstrate the applicability of our model, we conducted electronic structure calculations on a specific MOF crystal known as $Zn_4H_{28}C_{58}O_{13}$ (labeled as c6ce00407e_c6ce00407e6_clean) from the CoRE MOF database[61,62]. This particular crystal is one of the largest MOFs in the CoRE MOF database and possesses a large unit cell containing 5562 atoms. Using the universal HamGNN model, we obtain the energy bands and density of states for $Zn_4H_{28}C_{58}O_{13}$, as shown in Figure. 7. It can be seen from the energy bands that the structure has a bandgap value of about 2.6 eV, which makes it a suitable candidate for various applications in the field of energy harvesting and emission, such as photovoltaics and light-emitting diodes. Furthermore, we also predicted the difference charge density of this structure and observed that electrons primarily transfer from Zn and C atoms to the oxygen and hydrogen atoms. The difference charge density can visually display the distribution of static electric potential and possible catalytic sites.

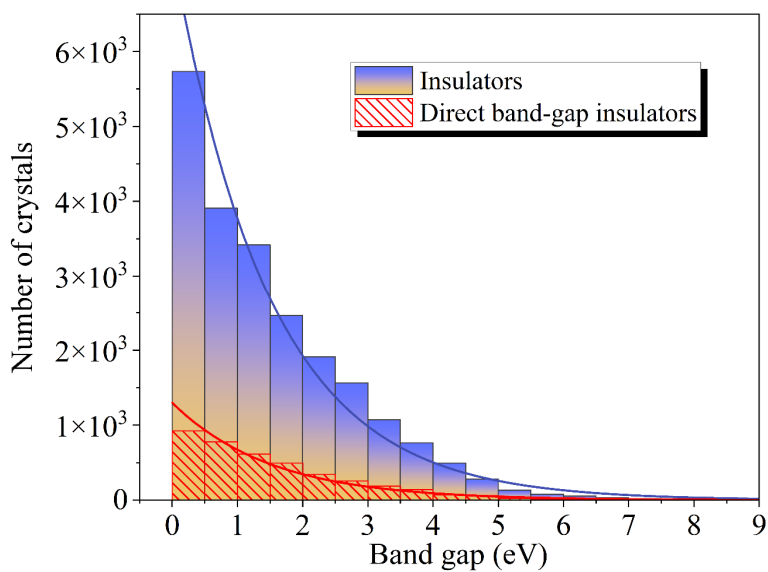

**Figure 8**. Histogram of the bandgap distribution for insulators and direct bandgap insulators.

## *High-throughput predictions of electronic structures for materials in the GeNOME dataset*

The GeNOME[63] database is a remarkable collection of 2.2 million stable crystal structures that have been discovered using large-scale active learning techniques. This dataset has expanded our knowledge of stable materials by almost an order of magnitude, providing an unprecedented opportunity for in-depth investigation of material properties and the development of novel machine learning models. Despite the stability of the GeNOME materials, their electronic structures remain largely unknown. While first-principles calculations could in principle determine the electronic structures, performing such computationally expensive calculations for the entire database would be prohibitively costly. Therefore, we employ the universal HamGNN model, which enables fast, accurate, and high-throughput predictions of electronic structures across the whole dataset, offering valuable insights for future theoretical predictions and experimental synthesis.

We employ the universal HamGNN model to compute the electronic structures of a total of 188,722 structures from the GeNOME database. Our analysis reveals that 21,973 of these structures exhibit insulating properties, with 3,940 being direct bandgap insulators and 18,033 being indirect band gap insulators. Direct bandgap materials have broad applications in photovoltaics and light-emitting devices due to their efficient electron-hole recombination processes. The histogram in Figure 8 illustrates the distribution of bandgaps for both insulators and direct bandgap insulators. The figure shows an exponential decline in the number of insulators as their bandgap values increase. Utilizing this universal Hamiltonian model not only enables the rapid selection of materials with direct band gaps but also facilitates the identification of crystals exhibiting flat bands near the Fermi level. This unique electronic structure holds great promise as it can give rise to intriguing properties and phenomena, such as fractional quantum Hall effects, Bose-Einstein condensation, unconventional superconductivity, and strong correlation effects[64-66]. After conducting a search using HamGNN, we have successfully identified 5109 crystals with flat bands. This number is nearly double compared to the 2379 flat-band crystals listed in the Materials Flatband Database. In order to demonstrate the accuracy of this approach for the GeNOME dataset, we will take a gapless material and a semiconductor material with flat bands, $Ti_6SeS_7$ and $K_3SrZr_6BI_{18}$, from the GeNOME database as examples. Notably, these two crystals are currently not included in the Materials Project.

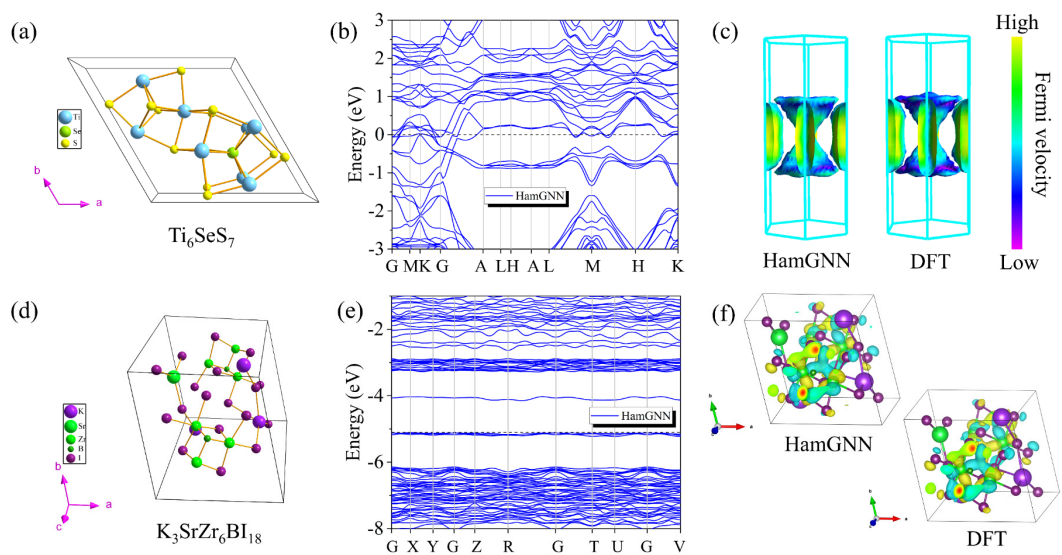

**Figure 9**. (a) The crystal structure of $Ti_6SeS_7$. (b) The predicted energy bands of $Ti_6SeS_7$ by the universal HamGNN model. (c) Comparison between the predicted Fermi surface of $Ti_6SeS_7$ and those calculated using DFT. (d) The predicted energy bands of $K_3SrZr_6BI_{18}$ by the universal HamGNN model. (e) The predicted energy bands of $K_3SrZr_6BI_{18}$ by the universal HamGNN model. (f) Comparison of the HamGNN predicted LUMO wave function at the G point and the DFT calculated LUMO wave function at the G point for $K_3SrZr_6BI_{18}$.

The prediction results for the electronic structures of $Ti_6SeS_7$ and $K_3SrZr_6BI_{18}$ are presented in Figure 9. In Figure 9(b), the predicted energy bands of $Ti_6SeS_7$ exhibit excellent agreement with the results obtained from DFT calculations, as shown in Supplementary Figure S10(a). In addition, we compare the Fermi surfaces obtained from our predicted Hamiltonian matrices with those computed using DFT in Figure 9(c). The comparison reveals remarkable similarities between the two Fermi surfaces, further validating the accuracy and reliability of our predictions. As shown in Figure 9(e), the presence of flat bands at both HOMO and LUMO levels in the predicted energy bands of $K_3SrZr_6BI_{18}$ is apparent. The predicted flat band electronic structure is validated by the DFT calculation (see Supplementary Figure. S10(b)). Figure 9(f) reveals that the flat band at the LUMO level is primarily formed by atomic orbitals from Zr and I atoms. This observation suggests a significant contribution from these elements to the unique electronic behavior exhibited by this compound. Furthermore, it is worth noting that our predicted wave function aligns well with actual calculated results, reinforcing the accuracy of our computational approach.

## Discussion

This work not only achieves a universal electronic Hamiltonian model for the entire periodic table but also demonstrates its practicality and reliability through successful predictions of electronic structures in various materials. We propose a framework to achieve a universal Hamiltonian model by training HamGNN on the crystal's Hamiltonian matrices obtained from the Materials Project or other large datasets. We have found that incorporating energy eigenvalues in the second training step, in addition to Hamiltonian matrices, is crucial for achieving a truly universal Hamiltonian model.

The universal model can accurately capture not only simple systems but also those containing uncommon or rare transition metal elements. One notable advantage of this universal model is its ability to handle complex multi-element systems comprising more than five elements. This capability opens up new possibilities for studying and understanding the electronic properties of advanced materials that often involve intricate combinations of chemical elements. The reliable framework provided by the universal electronic Hamiltonian model allows for efficient predictions of electronic structures across the periodic table. While the current publicly available GeNOME dataset contains 380,000 stable crystal structures, the complete dataset encompasses an unprecedented 2.2 million stable materials discovered through large-scale active learning.

We anticipate that with more structures being made publicly accessible, the Universal Hamiltonian model can be leveraged to rapidly and accurately compute the electronic band structures of this vast collection through high-throughput calculations with much less computational cost compared to first-principle calculations. This would enable the identification of a significantly larger number of crystals with desirable electronic properties, driving further advancements in materials science. This breakthrough has significant implications for material design, catalysis, electronics, and other fields that heavily rely on accurate knowledge and understanding of electronic properties.

## Methods

### *Network details.*

The equivariant node features are 32×0o+128×0e+128×1o+64×1e+128×2e+32×2o+64×3o+32×3e+32×4o+32×4e+16×5o+8×5e+8×6e, where '32×0o' means that there are 32 channels in this feature part, and the features in each channel are O(3) irreducible representations with $l = 0$ and odd parity. The node features utilized in this universal HamGNN model surpass those employed in our previous work[22] to enhance the network capacity for describing the entire periodic table. The universal HamGNN model has five orbital convolution layers. The spherical harmonic basis functions used to expand the interatomic directions are

0e + 1o + 2e + 3o + 4e + 5o + 6e. The interatomic distance between atom *i* and its neighboring atom *j*, which falls within the cutoff radius $r_c$, is expanded utilizing the Bessel basis function:

$$B(|\tau_{ij}|) = \sqrt{\frac{2}{r_c}} \frac{\sin(n\pi|\tau_{ij}|/r_c)}{|\tau_{ij}|} \quad (4)$$

The atomic neighbors are determined based on the cutoff radius of each atom's orbital basis. The interatomic distance is expanded using a series of Bessel functions with $n = 1, 2, \cdots, N_b$, where $N_b$ represents the number of Bessel basis functions. In this study, $N_b$ is set to 64.

### *DFT details.*

All the Hamiltonian matrices in the training set were computed using the PBE functional, with a Monkhorst-pack grid of 6×6×6, and a convergence criterion of $1.0 \times 10^{-8}$ Hartree. The energy cutoff employed for discretizing the real space is set at 200 Rydberg.

## Acknowledgment

We acknowledge financial support from the Ministry of Science and Technology of the People´s Republic of China (No. 2022YFA1402901), NSFC (grants No. 11825403, 11991061, 12188101), and the Guangdong Major Project of the Basic and Applied Basic Research (Future functional materials under extreme conditions--2021B0301030005).

## Data and code availability

The HamGNN code is publicly available from https://github.com/QuantumLab-ZY/HamGNN. The trained network weights for the universal model, the predicted energy bands for the test set, and the identified crystals with flat bands are available on Zenodo (https://zenodo.org/records/10827117).

## Author contributions

H.J.X. and X.G.G. supervised the project for the universal Hamiltonian framework. Y.Z. wrote the implementation codes of the universal Hamiltonian framework, and conducted network training and testing. X.Y.G. constructed the crystal structure of $Li_{10}Si_{1.5}P_{1.5}S_{11.5}Cl_{0.5}$ and discussed the corresponding findings. H.Y.Y. used the universal model to perform the high throughput calculation of electronic structures for the GeNOME dataset. J.H.Y. checked the results in the manuscript. Y.Z., H.Y.Y., H.J.X., and X.G.G. prepared the manuscript. All authors discussed the results and provided comments on the manuscript.

## Competing interests

The authors declare no competing interests.